# Binary galaxies and alternative physics.

## I. A qualitative application of MOND and Mannheim-Kazanas gravity

**D.S.L. Soares**

Departamento de Física, ICEX, UFMG, C.P. 702,

30.161-970 Belo Horizonte, MG

Brazil

E-mail: dsoares@fisica.ufmg.br

**Abstract.** Binary galaxies are modeled as point-masses obeying the non-Newtonian MOND and Mannheim-Kazanas (MKG) theories of gravity. Random samples of such systems are generated by means of Monte Carlo simulations of binary orbits. Model pairs have total masses and mass ratios similar to pairs in the cataloged sample used in the analysis.

General features of synthetic samples are derived from a comparison with observed data of galaxy pairs in $R \times \Delta V/(L_1+L_2)^{1/2}$ space. Both MOND and Mannheim-Kazanas binaries either on circular or low-eccentricity orbits cannot be the source of observations because they require extremely high $M/L$ values ($\approx 45$ solar units). Both MOND and MKG binaries on high-eccentricity orbits and reasonable $M/L$ values (5 solar units) produce envelopes of $R \times \Delta V/(L_1 + L_2)^{1/2}$ consistent with the observations, but the distribution of separations is inconsistent with the observed data, unless strong selection effects are at work.

A definite answer to the issue whether one or another model is suitable to explain real binary galaxy dynamics will be only possible when *a large sample containing a significant fraction of wide pairs, determined with velocity-blind selection procedures, is investigated*

astro-ph/9505014  19 Mar 96

Key words: binary galaxies – alternative physics – Monte Carlo simulations

## 1. Introduction

The dynamics pertaining to Newtonian gravity are unable of avoiding mass discrepancies in astronomical systems unless some kind of, yet unknown, dark particles are invoked (Faber and Gallagher 1979, van Albada and Sancisi 1986, Trimble 1987, Sanders 1990, Ashman 1992). For example, the MACHO and EROS searches for dark matter in the form of compact sub-luminous objects in our Galaxy halo (Alcock et al. 1993, Bennett et al. 1995, Aubourg et al. 1995, and references therein) have not detected sufficient microlensing events to explain the kinematics of the Milky Way with a spheroidal halo populated by brown dwarfs. Furthermore, direct optical searches of the Hubble Space Telescope have not also been able to detect the predicted amount of matter in the form of faint stars in the halo of the Milky Way (Bahcall 1994). Another much explored approach to the problem has been the suggestion that Newton laws are not adequate and that the solution might be found in alternatives to classical gravity and dynamical laws. Many of such alternatives are empirically motivated (e.g., Milgrom 1983a, 1983b, 1983c, Sanders 1984, Kuhn and Kruglyak 1987) by the observations of *flat rotation curves* but that is not always the case. Mannheim and Kazanas (1989, 1991) have derived an alternative theory of gravity whose main motivation was the search for the correct covariant general theory of gravity. Specifically, they found the exterior solution to conformal Weyl gravity associated with a static, spherically symmetric gravitational source.

The most successful of the empirical alternatives to Newton's laws has been Milgrom's MOND (*Modified Newtonian Dynamics*) which has been subjected to successive experimental tests (e.g., Gerhard 1994, but see Milgrom 1995). The theory has been strongly and successfully defended by Milgrom in many papers in recent years (see also Begeman, Broeils and Sanders 1991, Sanders 1994). Nevertheless, a most serious setback is that there is no covariant theory of gravity that, in the weak field limit, reduces itself to MOND. Some attempts have been made otherwise in this direction (Bekenstein and Milgrom 1984, Sanders 1986) but yet MOND remains as a phenomenological *ad hoc* theory, being this, up to now, its main source of criticism. Other consequences, dynamical and cosmological, of MOND are discussed by Felten (1984).

Mannheim (1992) has applied the Mannheim and Kazanas (1989) theory of gravity (hereafter MKG) to model the circular velocity profiles of four spiral galaxies with fairly reasonable fits, without the requirement of dark matter. He claims that Einstein-Newton gravity proves to be inadequate in comparison with fourth order conformal Weyl gravity, represented by the MKG solution. In spite of that, and of having on its foundation a general theory of gravity with the status of a full covariant one, MKG has been questioned on its ability of explaining exceptionally extended flat rotation curves (Sanders and Begeman 1994). Moreover, even conformal Weyl gravity has been criticized on cosmological grounds. Elizondo and Yepes (1994) presented exact solutions to the conformal Weyl gravity cosmological equations that failed to yield primordial nucleosynthesis abundances consistent with present observational constraints. They conclude that conformal cosmological models are very unlikely to give a realistic description of the Universe.

This is intended to be the first one of a series of papers devoted to binary galaxy dynamics, under the premises of alternative theories of gravity. Here, it is investigated what binary galaxy data might suggest one about MOND and MKG. Simple point-mass models are used to represent bound pairs of galaxies under MOND and MKG prescriptions for the gravitational mutual galaxy interaction. Monte Carlo simulations of synthetic samples are performed by means of numerical calculations of pair orbits. They are subsequently projected on the sky and the relevant observed quantities derived. A *qualitative* comparison between a sample of real pairs and simulated samples is then made. The second paper of the series will present a rigorous statistical analysis of observed samples of binaries in the light of MOND, and the third one will focus on MKG. Biases introduced by selection effects on sample determinations and "contamination" by non-physical pairs must be considered separately for each theory of gravity. It is obvious, for example, that the definition of what a non-physical pair is depends upon the theory that describes the interactions and the dynamics of galaxies in pairs.

Section 2 presents the data which come as a list of disk binary galaxies extracted from the *Catalogue of Multiple Galaxies*. A fiducial Keplerian model, determined in the Appendix, is qualitatively fitted to the data. Such a model is meant to be confronted with MOND and MKG simulated samples. In section 3, the MOND gravitational potential for point-mass binaries and technical details of the simulations are described. The same is done in section 4 for MKG. Section 5 is devoted to Monte Carlo simulations of MOND and MKG synthetic samples. In section 6, results of the qualitative comparison between observed and synthetic binaries are discussed and the main conclusions presented.

## 2. Binary galaxy data

Much effort has been dedicated in the last years to the compilation of binary galaxy lists (Karachentsev 1972, 1987, Turner 1976, Peterson 1979, Schweizer 1987, Soares 1989, Chengalur, Salpeter, Terzian 1993, 1994, Soares et al. 1995, etc). For the sake of uniformity, only one list of pairs is considered, namely, that extracted from the *Catalogue of Multiple Galaxies* (CMG, van Moorsel 1982, 1987, Oosterloo 1988, 1993, Soares 1989). The selection of pairs was done by applying a surface density enhancement procedure to the *Uppsala General Catalogue of Galaxies* (Nilson 1973), a method devised by T.S. van Albada. A description of the method can be found in van Moorsel (1982), Soares (1989, chapter 2, by van Albada and Soares) and Soares et al. (1995). The method does not require any redshift information, and this is a feature that makes it particularly suitable for investigations concerning alternative physics. The concept of a bound pair is dependent on what gravity law rules the dynamics of the pair, and cannot be relied upon simple upper limits on line-of-sight velocity differences, which is a typical characteristic of velocity-dependent selection procedures. Still, on uniformity, from the whole list of 358 binaries in the CMG a homogeneous subsample of 230 pairs is considered (see Soares 1989), which is formed by rotationally supported galaxies, ranging from lenticulars to late spirals. The main reason for avoiding pairs with elliptical galaxies is that both MOND and MKG have gathered their status of likely alternative theories after being applied on modeling circular velocity radial profiles of spiral galaxies. The whole sample is available on machine readable form upon request (`dsoares@fisica.ufmg.br`). The line-of-sight heliocentric velocity differences are normalized by the square root of the total blue luminosities of the pairs, in units of $10^{11} L_\odot$. Such a normalization allows one to scale the envelopes given by eqs. A11 and A14 (Figure 6 — Appendix) by a proper $M/L$ value.

It has been shown in many recent binary galaxy investigations, based on conventional physics, that there might be an extended dark halo either around the individual galaxies or embedding the whole pair (White et al. 1983, Schweizer 1987, Soares 1989, Charlton and Salpeter 1991, Bartlett and Charlton 1995). Although the presence of optical (i.e., non-physical) pairs in a sample could bias the analysis against point-mass models (as may have happened with the study by White et al. 1983, as pointed out by Pichio and Tanzella-Nitti 1985, and Pacheco and Junqueira 1988) there is compelling evidence that Keplerian models are far from being a good description of binary dynamics. Nevertheless, Keplerian approximations serve as reliable upper limits to $\Delta V/L^{1/2}$ for a sample of binary galaxies, provided one not be interested in the detailed distribution of pairs in that space.

In the Appendix, upper bounds on properly normalized velocity differences of galaxies in pairs as a function of projected separation in the plane of sky are derived. Such *envelopes* are here confronted with real data and set definite limits upon the overall distribution of the observations. It must be pointed out that the Keplerian model derived in the Appendix serves only as reference.

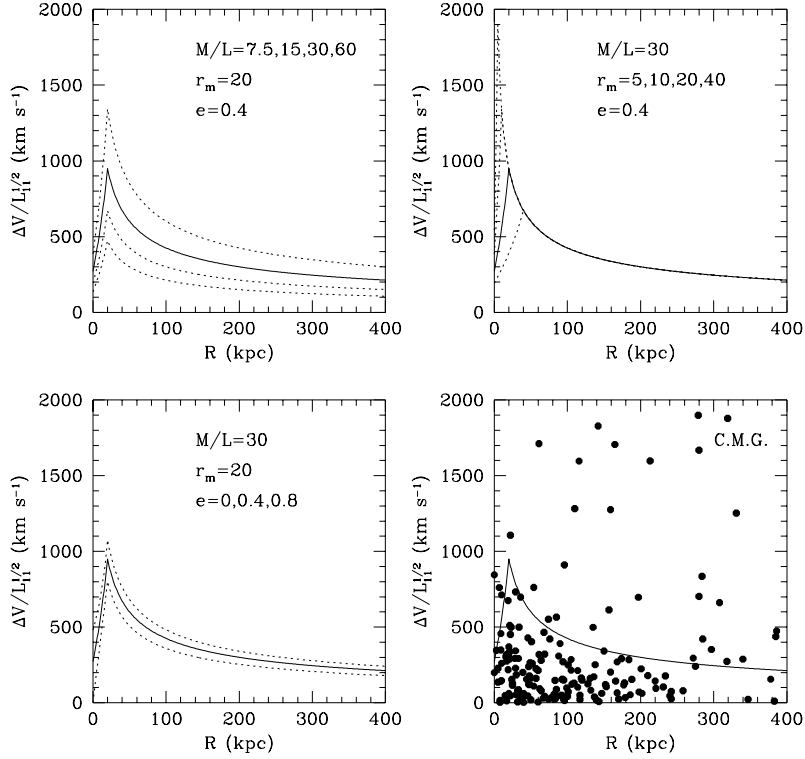

**Fig. 1.** Keplerian envelopes in the $R - \Delta V/L^{1/2}$ plane (curves). Filled circles represent CMG pairs. There are three model free parameters: the mass-to-light ratio ($M/L$), the minimum pericentric separation ($r_m$) and the orbital eccentricity ($e$). The solid curve, in the four panels, is the fiducial envelope mentioned in the text, with $M/L = 30$, $r_m = 20$ kpc and $e = 0.4$, chosen as a reference frame for CMG pairs $[L_{11} = (L_1 + L_2)/10^{11} L_\odot]$.

Figure 1 shows how to fill in the binary phase space with varying the three free parameters, $M/L$, $r_m$ (minimum pericentric separation) and $e$ (orbital eccentricity), allowed by the Keplerian envelope models derived in the Appendix. The ordinate $\Delta V/L^{1/2}$ is obtained by multiplying $\Delta V/M^{1/2}$ by the scaling factor $(M/L)^{1/2}$. The bottom right panel shows the CMG pair sample.

The solid curve in Fig. 1 was chosen as a fiducial envelope for the subsequent analysis. Its parameters are $M/L = 30$, $r_m = 20$ kpc and $e = 0.4$. From a Newtonian point of view, pairs above it are most likely unbound pairs, the so-called *optical pairs*, because they imply too high values of $M/L$. Incidentally, this is not necessarily true for MOND and MKG as it is shown in the Monte Carlo simulations below.

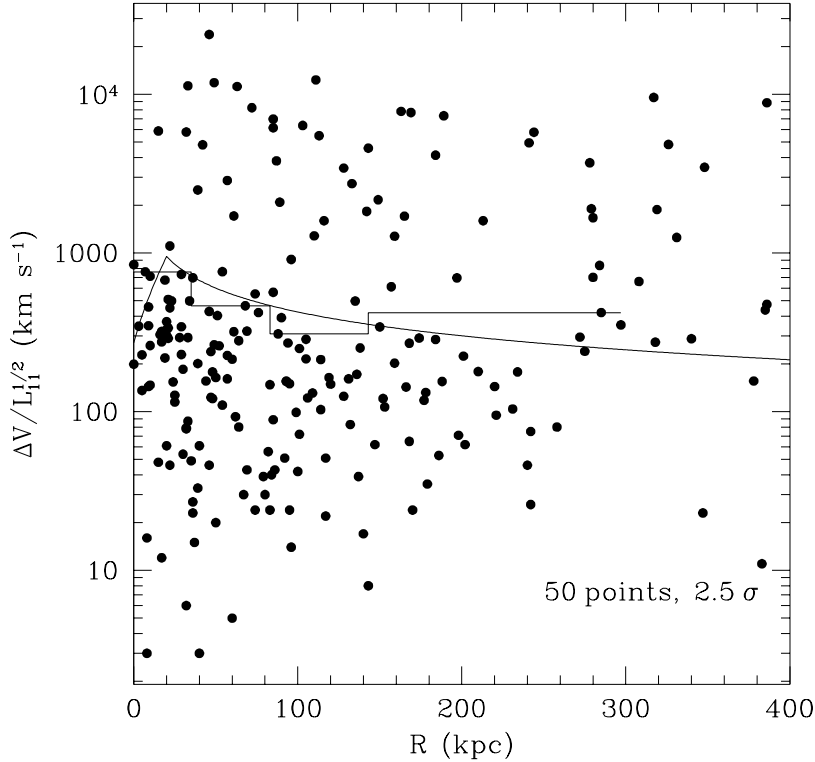

**Fig. 2.** The envelope of maximum $\Delta V/L^{1/2}$ is shown as a histogram, determined after exclusion of outliers. The number of pairs in each bin and the clipping weight are given in the right bottom corner of the panel. The smooth curve is the fiducial Keplerian envelope.

To verify whether such a Keplerian model is a fair reference envelope for the observations, a $k\sigma$ clipping procedure (see Yahil and Vidal 1977) was applied to the sample to exclude outliers, which would bias the determination of the envelope of maximum velocities. Figure 2 shows the resulting envelope confronted with the fiducial Keplerian envelope chosen above. Each bin in the histogram of Fig. 2 has 50 pairs. For every bin, the mean and standard deviation of $\Delta V/L^{1/2}$ are calculated, and pairs with $\Delta V/L^{1/2}$ over $2.5\sigma$ are excluded from the initial bin population. A new mean and $\sigma$ are calculated and the process is repeated. The iteration ends when every pair falls within $2.5\sigma$. Then, the highest $\Delta V/L^{1/2}$ is computed. It is apparent from Fig. 2 that the chosen parameters for the Keplerian envelope are consistent with the numerical envelope derived from the data.

## 3. MOND

The most successful phenomenological theory, alternative to dark matter Newtonian approach to the mass discrepancy problem, has been put forward by Milgrom, originally in three successive papers (1983a, 1983b, 1983c). It consists of a modification in Newtonian dynamics in the regime of low accelerations. In the framework of Milgrom's *Modified Newtonian Dynamics* (MOND), galaxies have no dark mass. Its simplified formulation has two basic assumptions: (i) Newtonian dynamics break down when the accelerations involved are small, and (ii) the acceleration $a$ of a test particle at a distance $r$ from a mass $m$ is given by $a^2/a_\circ \approx Gmr^{-2}$ in the limit $Gmr^{-2} \ll a_\circ$ (or $a \ll a_\circ$). The constant $a_\circ$ plays the role of a transition acceleration (implying a transition length scale, $r_t$, which is a function of the mass $m$) from the Newtonian to the MOND regime. In other words, in regions where the acceleration is smaller than the limit acceleration the gravitational field is roughly proportional to the inverse of the distance from the center of the galaxy, and as a consequence, spiral galaxy flat rotation curves can be obtained. The value of $a_\circ$ was originally determined by Milgrom (1983b) to be $\approx 4 \times 10^{-8}$cm/s$^2$ (A Hubble constant $H_\circ = 75$ kms$^{-1}$Mpc$^{-1}$ will be assumed throughout this paper). Later, a larger and better sample of flat rotation curves were analyzed and yielded the value of $1.21 \times 10^{-8}$cm/s$^2$, for the same $H_\circ$ (Kent 1987, Milgrom 1988 and Begeman, Broeils and Sanders 1991); such a value is used below in the present MOND simulations of binary galaxies.

The proper question here is in which way MOND works with binary galaxies. Milgrom (1986, 1994) gives numerical solutions of three problems in the framework of the Modified Newtonian Dynamics. They are: *(1)* the force law between two point masses, *(2)* rotation curves of various model disk galaxies, and *(3)* the field of a point mass in a constant external acceleration field. Of these, the first one is used here for the account of binary galaxy dynamics.

Milgrom finds, in problem *(1)*, that the MOND force field is very weakly dependent on the ratio $M_1/M_2 \equiv q$ of the two point masses. In the limit of large $r$ the force $F$ can be put in the form

$$F(M_1, M_2, r) = A(q) \times \frac{M_1 M_2}{(M_1 + M_2)^{1/2}} \frac{(Ga_\circ)^{1/2}}{r} \;, \tag{1}$$
$$A(q) = \frac{2}{3} q^{-1}(1+q)^{1/2} \left[(1+q)^{3/2} - q^{3/2} - 1\right].$$

This is the force field that one expects from a logarithmic potential. In the range $1 \le q \le 6$, characteristic of the CMG binary sample (see below), the dimensionless function

The test-particle circular velocity ($V_{\rm rot,i}$) of an individual galaxy, with mass equal to $M_i$, at distances larger than $r_t$, is

$$V_{\rm rot,i} = (GM_i a_{\rm o})^{1/4}, \qquad (2)$$

and the circular velocity [$V_{\rm circ}(r)$] of a binary system, with total mass equal to $M_1 + M_2$, is given by

$$V_{\rm circ}(r) = A(q)^{1/2}[G(M_1 + M_2)a_{\rm o}]^{1/4}. \qquad (3)$$

Equation 3 is valid in the limit of low accelerations, or, for binary galaxy separations much larger than the corresponding $r_t = (GM/a_{\rm o})^{1/2}$, where $M$ is the total binary mass $M_1 + M_2$. In the Monte Carlo simulations, described below, the total mass of pairs will range from $2.9 \times 10^{11} M_\odot$ to $1.3 \times 10^{12} M_\odot$. The transition radius for such systems ranges from 10 to about 20 kpc. Model galaxies have an assumed optical radius of about 10 kpc, thus, the binary potential is always MOND potential in the regime of $r \gg r_t$, as long as the visible galaxies do not overlap.

Each synthetic binary coming out of the simulation is obtained from the associated binary orbit. MOND orbits must be calculated numerically because the logarithmic potential does not admit an analytical solution. The starting point of the orbit calculation is the apocentric separation. Thus, one needs to know the apocentric separation $r_{\rm apo}$ itself and the velocity at apocenter. For a given eccentricity and apocentric separation, the velocity is univocally defined. It is convenient to express $V(r_{\rm apo})$ as a fraction of the circular velocity at apocenter, i.e., as $\alpha \times V_{\rm circ}(r_{\rm apo})$. In the simple case of a Keplerian orbit with eccentricity $e$, the parameter $\alpha$ is given by

$$\alpha(e) = (1 - e)^{1/2}. \qquad (4)$$

The relationship between $\alpha$ and the orbit eccentricity $e$, for the MOND binary potential, and galaxy separation $r \gg r_t$, is

$$\alpha(e) = \frac{(1-e)}{(2e)^{1/2}} \left[\ln\left(\frac{1+e}{1-e}\right)\right]^{1/2}, \qquad (5)$$

where $e$ is given by its general definition $e = (r_{\rm apo} - r_{\rm per})/(r_{\rm apo} + r_{\rm per})$. The function $\alpha(e)$, valid only in the low acceleration regime, is represented in Figure 3. An artificial sample of MOND binary galaxies is generated in section 5 by means of Monte Carlo simulations of binary orbits.

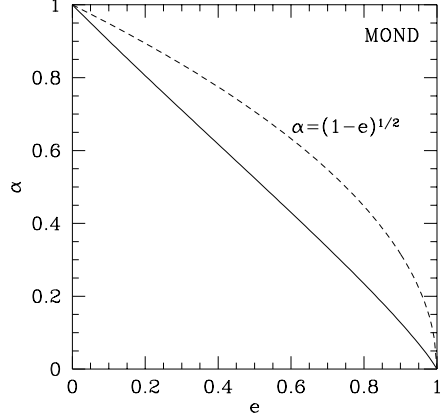

**Fig. 3.** Apocentric velocity in units of apocentric circular velocity as a function of orbital eccentricity for binary galaxy orbits. Solid curve: MOND potential, in the low acceleration regime (eq. 5). Dashed curve: Keplerian potential shown as a reference curve.

## 4. Mannheim-Kazanas gravity

Mannheim and Kazanas, in a series of papers (1989, 1991, 1994; Mannheim 1992, 1993, 1994a, 1994b) put forward the idea that gravity be based on the fourth order conformal Weyl theory rather than on Einstein's general relativity. While Einstein's gravity has a lot of success in a wide range of experimental facts, it faces its unequivocal inconsistency in the weak field limit, represented by Newtonian gravity, with the mass discrepancy problem in astronomical systems. Mannheim claims that one of the advantages of MKG conformal gravity is its ability, in the weak field limit, to solve the dark matter problem in spiral galaxies, explicitly posed by flat rotation curves. A conceptual point stressed by MKG's authors is that while alternative physics theories, such as MOND, depart from the phenomenological problem, which they are meant to explain (e.g., flat rotation curves), and then go on seeking a general covariant formulation, in order to get the desirable status of an acknowledged theory of gravity, Mannheim and Kazanas take the inverse way.

Instead of being based upon a second order action, as Einstein's general relativity, MKG follows from a conformal invariant fourth order action which implies a fourth order Poisson equation (Mannheim 1994b). The latter was integrated to yield an exact exterior solution to a spherically symmetric gravitational source. Neglecting terms similar to the standard general relativistic corrections to the Newtonian potential and a high order term in $r$ (see below), the following non-relativistic gravitational potential is derived,

$$U(r) = -\frac{Gm}{r} + \frac{\gamma c^2}{2}r = -\frac{Gm}{r}\left[1 - (\frac{r}{r_{\text{MKG}}})^2\right], \quad (6)$$

$$r_{\text{MKG}} = \left(\frac{2Gm}{\gamma c^2}\right)^{1/2} \approx 30(m/10^{11}M_\odot)^{1/2} \text{ kpc},$$

where $c$ is the speed of light and $\gamma$ a constant. The circular velocity of a binary system, with total mass $M_1 + M_2$, is then

$$V_{\text{circ}}(r) = \left[\frac{G(M_1 + M_2)}{r} + \frac{\gamma c^2}{2}r\right]^{1/2}. \quad (7)$$

The linear potential component in eq. 6 is responsible for the good fits obtained for flat rotation curves (e.g., Mannheim 1993). The actual value of $\gamma$, for a typical galaxy, is $\approx 10^{-28}$ cm$^{-1}$ (Mannheim and Kazanas 1989), which *"intriguingly is roughly the value of the inverse Hubble length"*, in the authors' words. In passing, Milgrom (1989) has also called the attention for *"one fact that may prove of prime importance"*, i.e., *"the near equality of $a_\circ$, [the MOND constant] as determined from galaxy dynamics, and that of $cH_\circ$, where $H_\circ$ is the present value of Hubble constant"*. Indeed, the value of $a_\circ$ adopted in this work (see section 3) is equal to 0.16 $cH_\circ$, and the value of $\gamma$ (see below) is 0.043 $H_\circ/c$, so one derives $a_\circ/\gamma = 3.7c^2$.

The reality of such relationships between $a_\circ$, $\gamma$ and $H_\circ$ would imply that the value of $H_\circ$ itself enters (non-linearly, see section 6) into the formalism of both MOND and MKG for interpreting binary orbits.

The MKG constant $\gamma$ is not an *universal* one, like MOND's $a_\circ$. It must be determined for every single galaxy, contrary to what is found with MOND (Begeman, Broeils and Sanders 1991). For the four galaxies studied by Mannheim (1993), $1/\gamma$ ranges from 1.3 to $4.0 \times 10^{29}$ cm. In the simulations of section 5, eq. 6 was used with the average $\gamma$, which has the inverse value of $2.5 \times 10^{29}$ cm. But the fact that $\gamma$ can vary from galaxy to galaxy must be taken into account when a sample of binaries with a substantial fraction of wide pairs is investigated. It must be pointed out however that it is not clear by which mechanism a law of gravity may have a tunable "constant" for every gravitational system. This seems to be a major weakness of MKG unless there is some constraint (of cosmological origin or of other kind) to fix $\gamma$ as a *universal* constant.

Orbits of point-masses under the potential given by eq. 6 are numerically calculated. The auxiliary parameter $\alpha$, described in the previous section, is related to the orbital eccentricity in a more complicated way than in Keplerian and MOND cases. The relationship includes dependences on the total binary mass $M_1 + M_2$ and on the orbital

$$\alpha(e) = \left\{ \left[\frac{1+x(e)}{x(e)^2}\right]\left[\frac{G(M_1+M_2)}{r_{\mathrm{apo}}} + \frac{\gamma c^2}{2}r_{\mathrm{apo}}\right] \right\}^{-1/2} \times$$

$$\times \left\{ 2\left[\frac{G(M_1+M_2)}{x(e)r_{\mathrm{apo}}} + \frac{\gamma c^2}{2}r_{\mathrm{apo}}\right] \right\}^{1/2}, \tag{8}$$

where $x(e) = (1-e)/(1+e) = r_{\mathrm{per}}/r_{\mathrm{apo}}$. Equation 8 is depicted in Figure 4 for $M_1+M_2 = 1 \times 10^{11} M_\odot$ and $r_{\mathrm{apo}} = 20, 50, 100$ and 400 kpc.

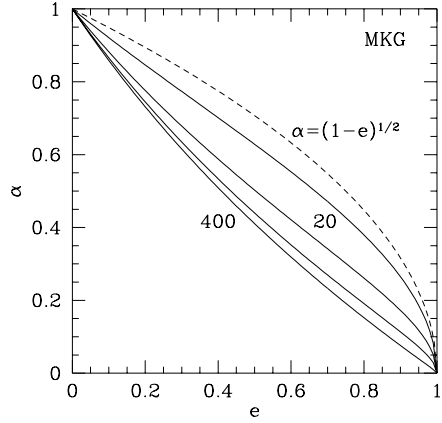

**Fig. 4.** Same as Figure 3 for the MKG potential (eq. 6). Solid curves: eq. 8 for $r_{\mathrm{apo}} = 20, 50, 100$ and 400 kpc, and $M_1+M_2 = 1 \times 10^{11} M_\odot$. Dashed curve: Keplerian potential shown as a reference curve.

In deriving eq. 6, some terms were neglected, including a higher order term in $r$. Mannheim and Kazanas (1989) and Mannheim (1995) speculate that on the largest scales, the contributions to such terms, and even to the linear component, from all galaxies, might merge and add to a preexistent general cosmological background. It is difficult to anticipate the gravitational influence of such a cosmological background, until a full development of MKG implications is available. It must be pointed out, however, that one deals here with *intermediate* scales ($\approx$ 1 Mpc and less), which are probably only little affected by the general cosmological version of eq. 6.

## 5. Monte Carlo simulations

The Monte Carlo simulations shown here follow the general recipe given in Soares (1990, hereafter S90), where a study of binary galaxies with dark halos was made. In that study all binaries had fixed $r_{\mathrm{apo}} = 200$ kpc. Now, for the sake of generality, a power-law distribution of binary spatial separations is:

$$P(r)dr \propto r^Q dr, \tag{9}$$

radius $r$ and external radius $r + dr$. The exponent $Q$, in the one-dimensional distribution $P(r)$, is approximately 0.2, if the two-point correlation function (which is usually expressed as a three-dimensional power-law distribution with power $-1.8$, e.g., Peebles 1980) is considered to be valid on scales of binary galaxies (Gott and Turner 1979, van Moorsel 1982, White et al. 1983). As in S90, a lower limit in $r$ of 20 kpc is set, dictated by the assumed size of a binary system at closest approach. The model binaries, both in MOND and MKG, have equal galaxies with 10 kpc of radius. An upper limit on $r$ of $\approx 1$ Mpc, although rather arbitrary, is also chosen. It is, otherwise, consistent with the widest pairs in the CMG sample and with recently determined wide pair lists (e.g., Chengalur et al. 1993). In other words, it is implicitly assumed that pairs of galaxies separated by more than $\approx 1$ Mpc are optical, be under the MOND or the MKG potential. The analysis made in the next section, however, is by no means affected by the choice of such an upper limit in $r$.

For technical reasons (see below), eq. 9 is applied to binary apocentric separations instead of spatial separations. There is no major problem with this procedure. In case of circular orbits, the distributions of spatial and apocentric separations are, of course, coincident, and, in case of non-circular orbits they are not significantly different due to the fact that orbiting bodies spend most of their orbital time near apocenter.

The simulated pairs need to have a distribution of total luminosities and luminosity ratios similar to the observed sample. The total masses of the simulated pairs are derived assuming $M/L = 5$, as found for MOND by Begeman, Broeils and Sanders (1991), and characteristic of stellar populations. Rather than adopting the distribution of observed $(L_1, L_2)$ for the simulated pairs, which requires computing 230 orbits, the simulated $(L_1, L_2)$ are obtained from the observed ones, averaged in a two-dimensional grid of $(q, L_1)$, where $q = L_1/L_2$. Table 1 shows this grid. 209 pairs (91% of the total) have $q$ less than 6. Mass ratios of the order of 10 or more are typical of satellite systems (e.g., Bontekoe and van Albada 1987), not representative for a study of binary dynamics, so they were avoided in the determination of the final 2D distribution $q \times L_1$. The grid shown in Table 1 has 24 cells and includes all pairs with $q \leq 6$, with 8 cells unoccupied.

The least massive binary system, $(6, 0.5)$, has a total mass of $2.9 \times 10^{11} M_\odot$ and the most massive one, $(3, 2.0)$, has a total mass of $1.3 \times 10^{12} M_\odot$. Such a two-dimensional grid defines the mass parameter space of artificial samples generated by Monte Carlo simulations. For the simulations of artificial pairs, the 16 occupied cells of Table 1 are sampled in proportion to the occupation frequency of the observed sample.

**Table 1.** Distribution of observed $q = L_1/L_2, L_1$

| q | $L_{1,11}$ =0.5 | $L_{1,11}$ =1.0 | $L_{1,11}$ =1.5 | $L_{1,11}$ =2.0 |
|---|---|---|---|---|
| 1 | 31 | 6 | 0 | 0 |
| 2 | 49 | 12 | 3 | 0 |
| 3 | 37 | 11 | 4 | 2 |
| 4 | 24 | 8 | 3 | 0 |
| 5 | 3 | 5 | 0 | 0 |
| 6 | 8 | 0 | 3 | 0 |

Note: $L_{1,11}$ is primary galaxy luminosity in units of $10^{11} L_\odot$.

Three different orbital eccentricities are considered, namely, $e = 0, 0.4$ and $0.9$, chosen as representative of low, medium and high eccentricity orbits. For every simulation panel shown in Figure 5, there are 2000 artificial pairs randomly distributed, as described above, throughout the 16 $(q, L_1)$ bins in Table 1, each pair being generated through the following steps. *(a)* A value of $r_{\rm apo}$ is obtained from the distribution given by eq. 9; *(b)* the value $\alpha = \alpha(e)$ is calculated (eq. 5 or eq. 8, for MOND and MKG, respectively); *(c)* $V(r_{\rm apo}) = \alpha \times V_{\rm circ}(r_{\rm apo})$ is calculated; *(d)* a test is done, using the conservation laws of energy and angular momentum, to verify if the pair will ever have a pericentric separation ($r_{\rm per}$) smaller than 20 kpc (the so-called *"merging test"*, adopted in S90). If this is true, the algorithm begins again at *(a)*. Otherwise, it goes to the next step. *(e)* A half-orbit is calculated, i.e., the binary path from apocenter to pericenter. The equations of motion, from MOND and MKG potentials, are numerically integrated. Due to the characteristics of the force field, this orbit segment is fully representative of the time evolution of the system; *(f)* the half-orbit is rotated by a random angle, in the orbital plane. This simple procedure mimics the evolution of the binary over a long time scale. *(g)* An orbital inclination $i$ is obtained according to the distribution $F(i) \propto \sin i$, i.e., the normal to the orbital plane is distributed at random in space. *(h)* Projected separation in the plane of sky ($R$), and line-of-sight velocity difference ($\Delta V$) are calculated, at a random time instant within the orbit half-period, as well as other relevant quantities, such as orbital eccentricity, orbital period, etc. This sequence is the same regardless of the gravitational potential adopted.

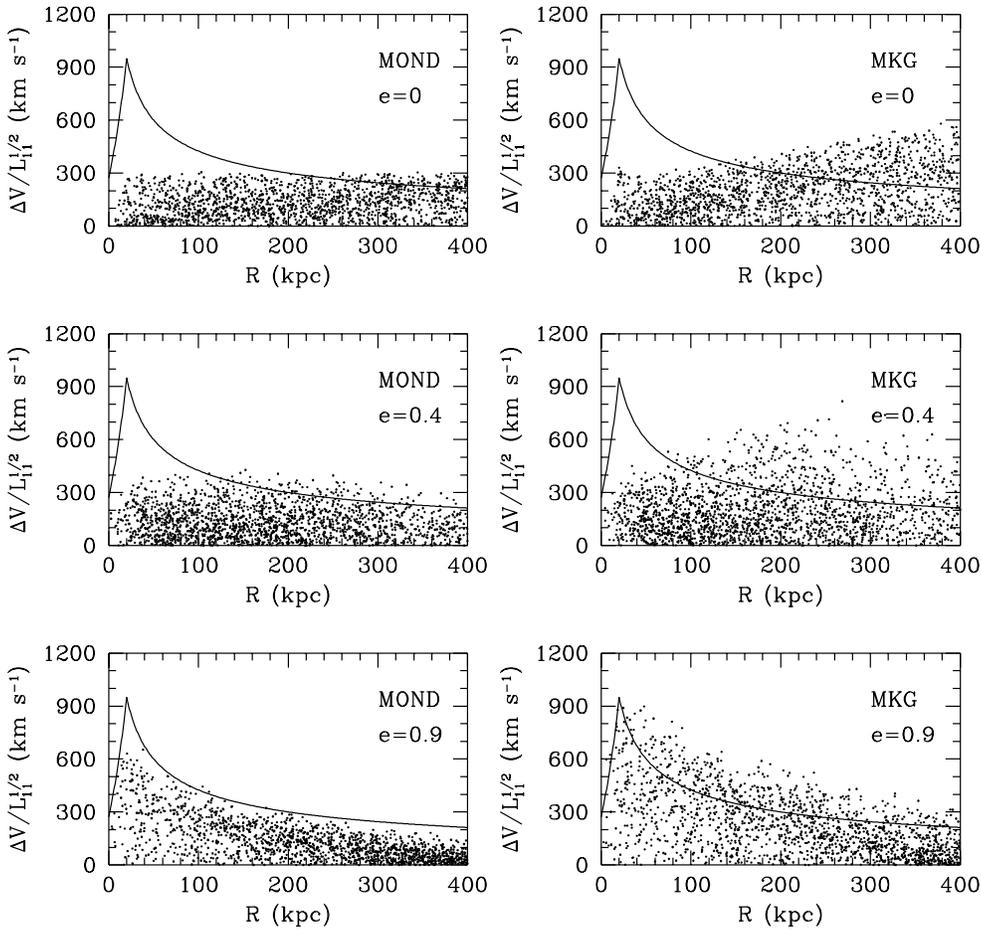

**Fig. 5.** Monte Carlo simulations of binary galaxies. Left panels represent binaries under MOND gravitational potential, and right panels under MKG gravitational potential. The solid curve is the fiducial envelope (Keplerian orbits, $M/L = 30$). The orbital eccentricities are indicated on the right upper corner of each panel, and $M/L = 5$ $[L_{11} = (L_1 + L_2)/10^{11} L_\odot]$.

## 6. Discussion and conclusions

Though it is not apparent from Fig. 5, it must be pointed out that, for a given $M/L$, the higher is the total mass of a given pair, the smaller is its $\Delta V/L^{1/2}$ ordinate. This does not happen with Keplerian models (see eqs. A11 and A14 ) but both MOND and MKG have approximately $\Delta V/L^{1/2} \propto M^{-1/2}$. A similar behavior was also seen with the dark halos models worked out in S90.

Figure 5 shows that MOND and MKG have distinct asymptotic behavior as $R$ increases. The upper envelope of the MOND circular orbit simulation is independent of $R$ while MKG's increases as $R^{1/2}$, as is expected from the linear potential. For non-circular orbits, MOND's upper envelopes have an asymptotic value given by $\alpha \times V_{\rm circ}(r_{\rm apo})/L^{1/2}$, and MKG's again increases with $R^{1/2}$. That is to say, the upper envelope of the MOND simulated points will eventually reach a plateau, for large $R$, and the MKG envelope will increase as $R^{1/2}$, since the $1/r$ component in the potential vanishes in that range of separations. The asymptotic limit for the Keplerian counterpart is of course zero. The asymptotic behavior of the various models is very promising as a way of discriminating amongst them, specially between MOND and MKG. The lack of a sufficiently large observed sample of wide pairs is, presently, the main obstacle to such an achievement.

From Fig. 5, in the range $R < 200$ kpc, it is readily clear that, for both MOND and MKG, binary galaxies in either *pure* circular orbits or in orbits with *only* intermediate eccentricity ($e = 0.4$) are not able of explaining the observations since they would require extremely high $M/L$ ($\approx 45$, in the range $R < 50$ kpc) for MOND and MKG standards. They also show different trends in the functional dependence of $\Delta V/L^{1/2}$ with $R$, which is apparent in the range $50 < R < 100$ kpc. *Pure* high eccentricity orbits ($e = 0.9$) represent viable solutions for both models. Elongated orbits are also consistent with the cosmological picture of binary galaxy formation in which such systems are formed by galaxies that instead of participating individually of the expanding universe stayed together because of their high mutual gravitational attraction, implying returning orbits with relatively low angular momentum (e.g., Schweizer 1987). A larger value of $a_{\rm o}$ would improve MOND's fitting since $\Delta V/L^{1/2} \propto a_{\rm o}^{1/4}$ (see eq. 3 for the low acceleration regime). In the case of MKG, on the contrary, a smaller value of $\gamma$ would do better since, in the large radius range where the linear potential component becomes significant, $\Delta V/L^{1/2}$ scales $\approx \gamma^{1/2}$ (see eq. 7), although here a smaller $M/L$, being still within MKG standards, would work as well on lowering the points on the $\Delta V/L^{1/2} - R$ plane. It is apparent that the only way to distinguish between MOND and MKG lies in the investigation of the certainly discriminate between MOND and MKG solutions.

Although their envelopes are consistent with the observed ones, high eccentricity orbits (MOND and MKG panels with $e = 0.9$) seem inconsistent with the observations because the corresponding simulations predict a large concentration of pairs at intermediate and large separations, as compared to small ones, and most of observed pairs have small separations. Nevertheless, this may be caused by selection effects. It is obvious that a combination of binary orbits with eccentricities ranging from small to large values is also possible for MOND and MKG. In this case, the high $\Delta V/L^{1/2}$ values (with $R < 200$ kpc) will be accounted for by high-eccentricity orbits while the low values will have contribution for the whole range of eccentricities. The way to discriminate between the two gravity models is an investigation of the relative frequency occupation of the $R - \Delta V/L^{1/2}$ plane. The large $R$ region (wide pairs), again, is fundamental in this aspect because of the very distinct predictions of MOND and MKG in that region (see Fig. 5). To accomplish such a discrimination, a full account of sample selection biases and of presence of optical pairs must be undertaken.

The above study gives one a secure indication of the input parameters for a more detailed statistical analysis, justifying the initial aim of the work, namely, to present an overview of specific applications of alternative theories of gravity to binary galaxy dynamics. The series beginning here will continue with a rigorous analysis of binary galaxy samples under the assumptions of MOND models. A third paper is planned on an investigation of binaries from the point of view of MKG. Both works will extend and complement the qualitative study presented here, with a careful account of selection effects on the determination of real samples of binary galaxies, with a quantitative measure of sample contamination by non-physical pairs, and with the introduction of the same effects in the list of prescriptions of Monte Carlo simulations of synthetic samples.

*Acknowledgements.* This work has been partially supported by the Brazilian *Conselho Nacional de Desenvolvimento Científico e Tecnológico, CNPq*, project number 300193/90-4. Drs. L.P.R. Vaz and G.A.P. Franco are thanked for reading the manuscript, for very important suggestions and for encouragement in the final stages of the work. L. Vertchenko is also thanked for an important suggestion. Dr. M. Milgrom is gratefully acknowledged for useful criticisms on an early version of this paper. The author would like to thank the paper's referee, Dr. G. Mamon: the paper has been considerably improved due to his suggestions.

## A. Keplerian models

The squared relative velocity of two orbiting point-masses $M_1$ and $M_2$ is:

$$V^2(r) = GM\left(\frac{2}{r} - \frac{1}{a}\right), \tag{A1}$$

where $G$ is the gravitational constant, $M = M_1 + M_2$, $r$ is the polar radial coordinate, and $a$ is the semi-major axis of the orbit. Consider now a two-dimensional phase space represented by the line-of-sight pair relative orbital velocity, normalized by the square root of the total pair mass, and the projected pair separation onto the plane of sky. These two quantities are related by the *projected* version of eq. A1:

$$R\left(\frac{\Delta V}{M^{1/2}}\right)^2 = \chi G, \tag{A2}$$

where $R$ is the pair projected separation and $\Delta V$ is the line-of-sight velocity difference. $\chi$ is a projection factor that takes care of the projection of both the orbital relative velocity and space separations of the galaxies in the pair:

$$\frac{\Delta V}{M^{1/2}} = \left[\frac{G}{a(1-e^2)}\right]^{1/2} \sin i [\cos(\nu - \omega) + e\cos\omega], \tag{A3}$$

and

$$R = \frac{a(1-e^2)}{1 + e\cos\nu}[1 - \sin^2(\nu - \omega)\sin^2 i]^{1/2}. \tag{A4}$$

The angles $\omega$, between the line of nodes and the orbit major axis, and $i$, the orbital plane inclination with respect to the plane of sky, give the orientation of the orbit in space. The true anomaly (polar angular coordinate) $\nu$ is the angle, in the orbital plane, between the line joining one galaxy to the other and the major axis of the orbit and $e$ is the orbital eccentricity. According to eqs. A3 and A4, the projection factor $\chi$ is given by:

$$\chi = \frac{\sin^2 i}{1 + e\cos\nu}[\cos(\nu - \omega) + e\cos\omega]^2 \times$$
$$\times [1 - \sin^2(\nu - \omega)\sin^2 i]^{1/2}. \tag{A5}$$

Note that, for a given eccentricity, $\chi$ has a maximum value equal to $\chi_{\max}(e) = 1 + e$ (the conditions for a maximum projection are $\omega = 0, i = \pi/2$ and $\nu = 0$. An immediate conclusion from eq. A2 is that any observed sample of pure bound pairs would appear below the curve

$$R\left(\frac{\Delta V}{M^{1/2}}\right)^2 = \chi_{\max}(1)G = 2G \tag{A6}$$

in a $\Delta V/M^{1/2} - R$ diagram.

that as $R$ approaches zero, $\Delta V/M^{1/2}$ becomes increasingly larger. This would be true if galaxies were point-masses, which would imply no restriction whatsoever in orbital sizes, that is to say, the major axes could have any value. Of course, this is not possible with real galaxies; thus, one needs to put a lower limit on orbital sizes. This limit can be defined in terms of a minimum allowed pericentric distance, $r_{\text{per}} \equiv r_m$. This implies in a minimum semi-major axis, $a_{\min}(e) = r_m/(1-e)$, for a Keplerian orbit with eccentricity $e$. In fact, such condition is equivalent to saying that pairs with pericentric separations smaller than $r_m$ will merge quickly, which is tantamount to introducing a *merging condition* on Keplerian pairs.

Upper envelopes, similar to eq. A6, are now derived considering several orbital eccentricities, and the pericentric cutoff described above. Orbits that have $a = a_{\min}$ lead to maximum $\Delta V/M^{1/2}$ :

$$\frac{\Delta V}{M^{1/2}} = \left[\frac{G}{r_m(1+e)}\right]^{1/2} \sin i [\cos(\nu - \omega) + e \cos \omega], \tag{A7}$$

at

$$R = \frac{r_m(1+e)}{(1+e \cos \nu)}[1 - \sin^2(\nu - \omega) \sin^2 i]^{1/2}. \tag{A8}$$

Substituting the angular conditions for maximum projection ($i = \pi/2, \omega = 0$ and $\nu = 0$) one gets

$$\left(\frac{\Delta V}{M^{1/2}}\right)_{\max} = \left[\frac{G(1+e)}{r_m}\right]^{1/2} \tag{A9}$$

that occurs at

$$R\left[\left(\frac{\Delta V}{M^{1/2}}\right)_{\max}\right] = r_m. \tag{A10}$$

For a given eccentricity there is a well-defined maximum $\Delta V/M^{1/2}$; for all eccentricities, maxima occur at the same $R = r_m$. In the case of $R \geq r_m$ the curve that describes the maximum values of $\Delta V/M^{1/2}$ is

$$\frac{\Delta V}{M^{1/2}}(R, e) = \left[\frac{G(1+e)}{R}\right]^{1/2}. \tag{A11}$$

and corresponds to orbits with fixed $e, a = a_{\min}(e) \geq a_{\min}(0) = r_m, i = \pi/2, \omega = 0$ and $\nu = 0$. In the range $0 \leq R < r_m$, one finds the largest possible values of $\Delta V/M^{1/2}$ by fixing $a = a_{\min}, i = \pi/2$, and $\omega = 0$, but letting $\nu$ (that gives the relative position of galaxies in orbit) be a free parameter. Doing that in eqs. A7 and A8 one gets:

$$\frac{\Delta V}{M^{1/2}} = \left[\frac{G}{r_m(1+e)}\right]^{1/2}(\cos \nu + e) \tag{A12}$$

$$R = \frac{r_m(1+e)\cos\nu}{1+e\cos\nu}. \tag{A13}$$

Eliminating $\nu$ in eqs. A12 and A13 yields the equation of the inner maximum envelope:

$$\frac{\Delta V}{M^{1/2}}(R,e) = \left[\frac{G}{r_m(1+e)}\right]^{1/2}\left[\frac{R}{r_m(1+e)-eR}+e\right]. \tag{A14}$$

Note that, for $e = 0$ (circular orbits), eq. A14 reduces to a line, $\Delta V/M^{1/2} = (G/r_m^3)^{1/2}R$, and, for $R = r_m$, i.e., the position of the maximum value of $\Delta V/M^{1/2}$, it takes the form of eq. A9, as it should. Figure 6 shows the envelopes for various eccentricities, and for the total pair mass given in units of $10^{11}M_\odot$. The most important point to notice is that the merging condition, introduced by only having $a \geq a_{\min}$, does not allow pairs with high $\Delta V M^{1/2}$ at small separations (see dashed curves in Fig. 6).

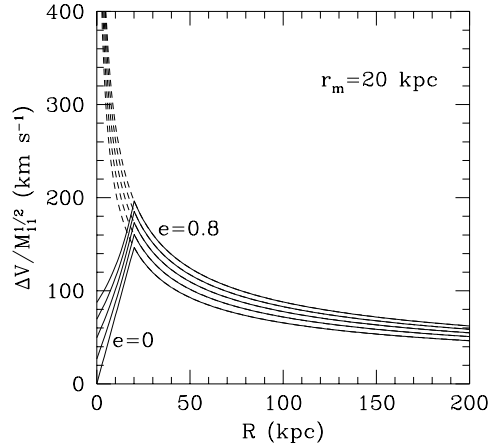

**Fig. 6.** The maximum allowed values of $\Delta V/M^{1/2}$, as a function of $R$ and eccentricity $e$, are represented by the solid curves (eq. A11, for $R \geq r_m$, and eq. A14, for $R < r_m$). The dashed curves show the envelopes for $R < r_m$, if galaxies are point-masses without restrictions on the pericentric separation. The values of the eccentricities are 0, 0.2, 0.4, 0.6 and 0.8. The total pair mass is given by $M_{11} = (M_1 + M_2)/10^{11} M_\odot$.

### References

van Albada T.S., Sancisi, R., 1986, Phil. Trans. R. Soc. London, Ser. A 320, 447

Alcock C. et al., 1993, Nat 365, 621

Ashman K.M., 1992, PASP 104, 1109

Aubourg E. et al., 1995, A&A 301, 1

Bahcall J., 1994, StScI release no. PRC94-41a